\documentclass[12pt,epsf,amstex]{article}
\usepackage [dvips]{graphicx}
\usepackage{amsmath}
\usepackage{amssymb}
\usepackage{epsfig}
\usepackage{verbatim}
\usepackage{color}

\addtocounter{secnumdepth}{1}
\usepackage[utf8]{inputenc}   
\usepackage[T1]{fontenc}
\usepackage{ae,aecompl}

\begin{document}
\newcommand{\ab}{b}
\newcommand{\bE}{\bar{E}}
\newcommand{\calC}{{\cal C}}
\newcommand{\calH}{{\cal H}}
\newcommand{\calJ}{{\cal J}}
\newcommand{\calM}{{\cal M}}
\newcommand{\calN}{{\cal N}}
\newcommand{\hP}{\hat{P}}
\newcommand{\hPi}{\hat{\Pi}}
\newcommand{\sumn}{\sum_{n=1}^N}

\newcommand{\bnu}{\bar{\nu}}
\newcommand{\bnuone}{\bar{\nu}_1}
\newcommand{\bnutwo}{\bar{\nu}_2}

\newcommand{\veca}{\vec{a}}
\newcommand{\vecA}{\vec{A}}
\newcommand{\ai}{{a_i}}
\newcommand{\aone}{{a_1}}
\newcommand{\atwo}{{a_2}}
\newcommand{\Ai}{{A_i}}
\newcommand{\Aone}{{A_1}}
\newcommand{\Atwo}{{A_2}}

\newcommand{\vecalpha}{\vec{\alpha}}
\newcommand{\vecg}{\vec{g}}
\newcommand{\vecp}{\vec{p}}

\newcommand{\tA}{\tilde{A}}
\newcommand{\tB}{\tilde{B}}
\newcommand{\tP}{\tilde{P}}
\newcommand{\tbeta}{\tilde{\beta}}
\newcommand{\tgamma}{\tilde{\gamma}}
\newcommand{\tcalM}{\widetilde{\cal M}}
\newcommand{\betast}{{\beta_*}}
\newcommand{\fstar}{f_*}

\newcommand{\intp}{\int_{-\pi}^{\pi}\frac{\dd p}{2\pi}}
\newcommand{\intpone}{\int_{-\pi}^{\pi}\frac{\dd p_1}{2\pi}}
\newcommand{\intptwo}{\int_{-\pi}^{\pi}\frac{\dd p_2}{2\pi}}
\newcommand{\ointz}{\oint\frac{\dd z}{2\pi{\rm i}}}
\newcommand{\qext}{q_{\rm ext}}

\newcommand{\bO}{{\bf{O}}}
\newcommand{\bR}{{\bf{R}}}
\newcommand{\bS}{{\bf{S}}}
\newcommand{\bT}{\mbox{\bf T}}
\newcommand{\bt}{\mbox{\bf t}}
\newcommand{\half}{\frac{1}{2}}
\newcommand{\thalf}{\tfrac{1}{2}}
\newcommand{\bsA}{\mathbf{A}}
\newcommand{\bsV}{\mathbf{V}}
\newcommand{\bsE}{\mathbf{E}}
\newcommand{\bsT}{\mathbf{T}}
\newcommand{\bsZ}{\hat{\mathbf{Z}}}
\newcommand{\bse}{\mbox{\bf{1}}}

\newcommand{\invup}{\rule{0ex}{2ex}}

\newcommand{\bGamma}{\boldmath$\Gamma$\unboldmath}
\newcommand{\dd}{\mbox{d}}
\newcommand{\ee}{\mbox{e}}
\newcommand{\p}{\partial}

\newcommand{\cdottt}{\!\cdot\!}
\newcommand{\wt}{\widetilde}
\newcommand{\wh}{\widehat}

\newcommand{\parity}{{\mbox{par}}}

\newcommand{\beq}{\begin{equation}}
\newcommand{\eeq}{\end{equation}}
\newcommand{\bea}{\begin{eqnarray}}
\newcommand{\eea}{\end{eqnarray}}
\def\lsim{\:\raisebox{-0.5ex}{$\stackrel{\textstyle<}{\sim}$}\:}
\def\gsim{\:\raisebox{-0.5ex}{$\stackrel{\textstyle>}{\sim}$}\:}

\thispagestyle{empty}
\title{\Large {\bf On the oddness of percolation}              \\[2mm] 
}

\author{{C.~Appert-Rolland and H.J.~Hilhorst}\\[5mm]
{\small Laboratoire de Physique Th\'eorique (UMR 8627)}\\[-1mm] 
{\small CNRS, Universit\'e Paris-Sud}\\
{\small Universit\'e Paris-Saclay, 91405 Orsay Cedex, France}\\}

\maketitle
\begin{small}
\begin{abstract}
Recently Mertens and Moore [{\it Phys.~Rev.~Lett.} {\bf 123} (2019) 230605] 
showed that site percolation ``is odd.''
By this they mean that on an $M\times N$ square lattice the number of distinct
site configurations that allow for vertical percolation is odd.
We report here an alternative proof,
based on recursive use of geometric symmetry, for both free and periodic
boundary conditions.

\end{abstract}
\end{small}
\vspace{12mm}

\newpage 

\begin{figure}[ht]
\begin{center}
\scalebox{.45}
{\includegraphics{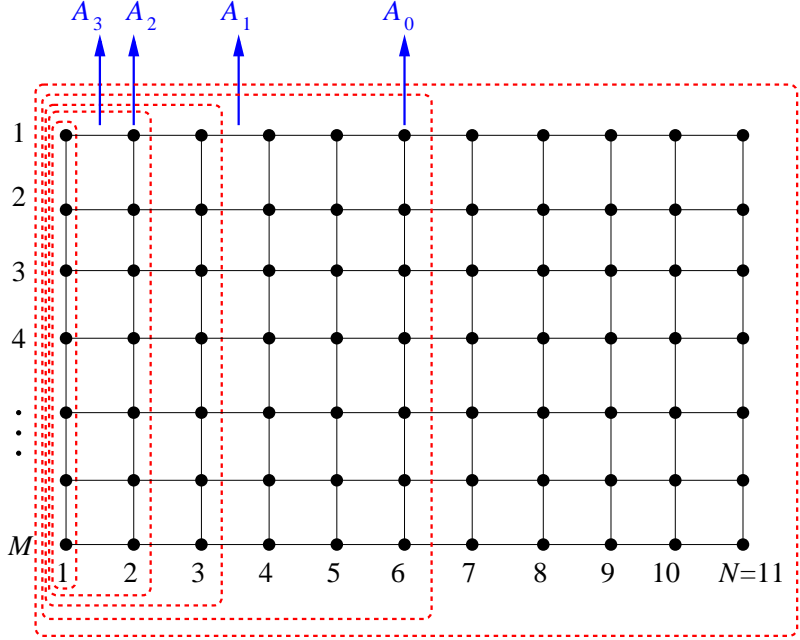}}
\end{center}
\caption{\small 
Square lattice of $M\times N$ sites with nearest-neighbor links.
Each lattice site may be ``open'' or ``closed'' (not shown).
The proof in the text makes use of
a nested sequence of sublattices of sizes $M\times N_j$, for
$j=0,1,\ldots,j_{\rm max}$. In this example each sublattice is identified by 
its surrounding dashed red contour; we have
$N=N_0=11, N_1=6, N_2=3, N_3=2, N_4=1$, and $j_{\rm max}=4$.
The arrow $A_j$ indicates the central vertical axis
of the $M\times N_j$ sublattice.
For any $N$ the innermost sublattice is a single column of $M$ sites. 
}
\label{fig_one}
\end{figure}

Models of percolation have been among the
most studied lattice problems in statistical mechanics
\cite{BroadbentHammersley57,Essam80} ever since the 1950s.
Surprisingly, even today new results appear.
Mertens and Moore \cite{MertensMoore19} very 
recently considered site percolation
on a square lattice of $M$ rows and $N$ columns ($M,N\geq 1$)
as shown in figure \ref{fig_one}.
Each site may be open or closed and therefore there are $2^{MN}$ distinct 
site configurations.
The lattice has free boundary conditions.
A vertically percolating configuration is one that has an open path 
joining the top row and the bottom row 
along nearest-neighbor links.
The authors of Ref.\,\cite{MertensMoore19} prove the oddness of
the total number of percolating configurations. 
Their method of proof begins by
distinguishing between the class of percolating configurations
for which the total number of open sites
is even and the class for which it is odd;
after pairwise elimination of configurations from both classes
their proof ends with a solvable combinatorial problem.

Here we present an alternative proof of the same result.
Our proof does not refer to the two classes distinguished in 
Ref.\,\cite{MertensMoore19}.
It relies on a recursive pair elimination process, 
governed by reflection of the site configurations with respect to
a sequence of vertical axes;
at the end of the elimination process only a single configuration is left,
which immediately implies the ``oddness of percolation.''
We believe this alternative proof also merits to be reported
and we do so below.
Finally we extend the proof
to other types of boundary conditions.
\\

Let $\calC_{MN}$ be the subset of all $2^{MN}$ configurations that
percolate vertically under free boundary conditions. 
We focus directly on the number 
$|{\calC_{MN}}|$ of its elements.
The statement to be proven is that
the parity $\parity(|{\calC_{MN}}|)$ is odd.

For an arbitrary configuration $\sigma\in\calC_{MN}$ we consider its
reflection about the central vertical axis of the lattice, 
{\it i.e.,} we exchange its columns $n$ and $N+1-n$ for 
all $n=1,2,\ldots,\lfloor{\tfrac{N}{2}}\rfloor$.
If $N$ is odd, then the central axis passes through column 
$\tfrac{N+1}{2}$ and this column is invariant; if $N$ is even, then 
the central axis passes in between columns $\frac{N}{2}$ and
$\tfrac{N}{2}+1$. 
We let $\sigma^\prime$ denote the reflected configuration.

If $\sigma\neq\sigma^\prime$, then we may eliminate $\sigma$ and $\sigma^\prime$
simultaneously from set
$\calC_{MN}$ without affecting its parity.
Going through all configurations $\sigma\in\calC_{MN}$
and eliminating pairs whenever applicable,
we obtain a reduced set $\calC^{\rm sym}_{MN}$ of configurations
that are all left-right symmetric ($\sigma=\sigma^\prime$), and this set is 
such that $\parity(|\calC^{\rm sym}_{MN}|)=\parity(|\calC_{MN}|)$.

Any configuration $\sigma\in\calC^{\rm sym}_{MN}$, 
because of this symmetry property,
may be represented by its ``left half,'' that is, by its site configuration 
restricted to the sublattice consisting of only columns
$1,2,\ldots,N_1$, where $N_1=\lceil{\frac{N}{2}}\rceil$ 
(see figure \ref{fig_one}).
The fact that $\sigma$ percolates and is symmetric implies that
its configuration restricted to the sublattice of $N_1$ columns also
percolates and therefore this restricted configuration is in $\calC_{MN_1}$.
Inversely, under free boundary conditions
every member of $\calC_{MN_1}$ has a unique symmetric extension
that is in $\calC^{\rm sym}_{MN}$.
Hence the elements of $\calC^{\rm sym}_{MN}$ are in one-to-one
correspondence with those of $\calC_{MN_1}$.
It follows that 
$\parity(|\calC_{MN}|)=\parity(|\calC_{MN_1}|)$, and we have thereby transformed
the original problem on an $M\times N$ lattice into an identical one on a
sublattice of reduced size $M\times N_1$.

We now iterate this reduction as illustrated in figure \ref{fig_one}.
In the $j$th iteration step we exploit
reflection about the central vertical axis of
the $M\times N_{j-1}$ sublattice.
This leads to the recursion 
\beq
\parity(|\calC_{MN_{j}}|) = \parity(|\calC_{MN_{j-1}}|)\,, \qquad 
N_{j} = \Bigg\lceil{\frac{N_{j-1}}{2}}\Bigg\rceil,
\label{piiter}
\eeq
valid for $j=1,2,\ldots,j_{\rm max}$
with initial condition $N_0=N$,
and where $j_{\rm max}$ is the first value of $j$ such that $N_j=1$. 
Therefore
\beq
\parity(|\calC_{MN}|) = \parity(|\calC_{M1}|).
\label{pi1}
\eeq
The right-hand side of this identity refers to a lattice consisting of 
a single column of height $M$.
The set $\calC_{M1}$ of all 
percolating site configurations
on this column contains only a single member, 
namely, the column with all sites open 
(the corresponding reconstructed element of $\calC_{MN}$ is the 
configuration in which all $MN$ sites are open).
Therefore 
$\parity(|{\calC_{M1}|)} = \parity(1) = \mbox{odd}$,
which when inserted in (\ref{pi1}) yields
\beq
\parity(|\calC_{MN}|) = \mbox{odd}.
\label{pimnisone}
\eeq
This completes the oddness proof of $|\calC_{MN}|$.
It has relied on nothing but
the successive pairwise elimination of 
percolating site configurations,
according to their symmetry properties under reflection
with respect to appropriately chosen vertical axes.
In the end only a single configuration remains, namely the
fully open lattice.
The column height $M$ appears as a mere parameter
and plays no role in the proof.\\

We briefly consider vertical percolation for other boundary conditions,
allowing these now to be free (F) or periodic (P) in either direction.
This leads to the four cases FF, FP, PF, and PP,
where the first and second letter refer to the $x$ and $y$ direction, respectively.  The proof presented above is for the FF case.
It will appear that in all other cases, too, percolation is odd.

For FP boundary conditions
the lattice is a cylinder of length $N$ and circumference $M$.
A configuration is said to be vertically percolating if it
contains an open path that wraps around the cylinder.
The initial subset $\calC_{MN}$ consists of all configurations that 
percolate vertically in this new sense.
The central axis $A_0$ 
defines two sublattices of size $M\times\lceil\frac{N}{2}\rceil$ that are also 
cylinders. Configurations that are not symmetric about this axis 
occur in pairs and are
eliminated as before; and the remaining 
symmetric configurations on the original lattice are in one-to-one correspondence with the 
vertically percolating configurations on one of the sublattices.
Hence the problem with parameter $N$ reduces to an identical problem with length $\lceil\frac{N}{2}\rceil$. Iteration leads to
a single column of sites with periodic boundary conditions
in the vertical direction, that is, a ring.
Only a single configuration percolates, namely the one with all sites open,
and hence vertical percolation under FP boundary conditions is odd.

For PF boundary conditions
the lattice is a cylinder of height $M$ and circumference $N$.
The iteration procedure now requires an exceptional first step.
We choose a vertical axis $\bar{A}_0$ passing midway between the columns
$N$ and $1$ (whose sites are now connected by links) and another vertical axis $A_0$ diagonally opposite to $\bar{A}_0$.
This second axis will be midway between two columns (for $N$ even) or
pass through the sites of a column (for $N$ odd) exactly as in the FF case. 
We note that configurations symmetric under reflection about $A_0$
are also symmetric under reflection with respect to $\bar{A}_0$. 
Among all the vertically percolating configurations,
the nonsymmetric ones are pairwise eliminated.
For each remaining symmetric configuration, its restriction to
one of the two sublattices delimited by $\bar{A}_0$ and $A_0$ also percolates.
Thus the symmetric configurations on the cylinder
are in one-to-one correspondence with the 
vertically percolating configurations on an $M\times\lceil\frac{N}{2}\rceil$ lattice with
FF boundary conditions. This reduces the problem to the FF case
and hence vertical percolation under PF boundary conditions is odd.

For PP boundary conditions we deform the lattice to a torus
whose major radius corresponds to the lattice periodicity in the $x$ direction. The lattice columns therefore become small circles 
(circles with minor radius). 

Again, an exceptional first step is needed.
Let the torus be divided into two symmetric halves by a vertical plane that passes through its axis and that we orient such that it cuts
the torus midway between the two small circles at positions $N$ and $1$. The cut on the diagonally opposite side will either be midway between two small circles (for $N$ even)
or coincide with one such small circle (for $N$ odd).
These two cuts are the obvious analogs of the axes  $\bar{A}_0$ and ${A}_0$  of the PF case.
Each half of the torus so obtained is a cylindrical sublattice of length
$\lceil\frac{N}{2}\rceil$ and of circumference $M$.
Configurations which are nonsymmetric with respect to the dividing plane are pairwise eliminated.
The remaining symmetric configurations on the torus are in one-to-one correspondence with the vertically wrapping configurations on one of the two cylindrical sublattices. This reduces the problem to the FP case
and hence vertical percolation under PP boundary conditions, that is, on the torus, is odd.
We note that the initial set of vertically percolating configurations on the torus includes the topologically special
subset where the open path spirals. Since such configurations cannot be symmetric under reflection about the dividing plane,
they are eliminated by the first special step and play no special role.\\

We will say that a configuration {\it bipercolates\,} if it percolates in both the $x$ and the $y$ direction. 
Bipercolation has been considered 
in the context of conformal field theory \cite{Pinson94,Watts96}.
Our method is easily extended to show that the parity of the set of all bipercolating configurations on an $M\times N$ lattice is odd under all
boundary conditions considered above.
The essential observation is that whenever a configurations in the initial set $\calC_{MN}$
percolates horizontally in addition to its vertical percolation,
then the reduced configuration obtained by reflection with respect to vertical axes also percolates horizontally.\\

The strategy applied above, where {\it vertical\,} percolation is analyzed
in terms of reflections about {\it vertical\,} axes, is not unique.
For FF and PF boundary conditions one
may also consider reflections about a sequence
of {\it horizontal\,} axes and stepwise reduce the number
of rows to only a single one. That row, of length $N$, has $2^N$
distinct site configurations all of which percolate vertically
except for the one with all $N$ sites closed. 
That is, the number of percolating configurations is $2^N-1$, which is
odd, and hence $\parity(|\calC_{MN}|)$ is odd.\\

Eq.\,(\ref{pimnisone}) was originally established by 
Mertens and Moore \cite{MertensMoore19} for the cases FF and PF
(no wrapping).
These authors subsequently obtain many interesting additional
results for more general lattices and boundary conditions.
It is not within our scope here 
to investigate extensions of 
our method of proof to such settings.\\

{\bf Acknowledgment.}
We thank one of the referees for interesting questions that have led us
to extend the discussion of boundary conditions.


\appendix

\end{document}